\documentclass[letterpaper,aip,jcp, floatfix,
reprint
]{revtex4-1}

\usepackage{amsmath}
\usepackage{graphicx}
\usepackage{dcolumn}
\usepackage{bm}
\usepackage[caption=false]{subfig}

\begin{document}

\preprint{}

\title{Characterization of a Vacuum Ultraviolet Light Source at 118~nm}



\author{John M. Gray }
\author{Jason Bossert}
\author{Yomay Shyur}
\author{Ben Saarel}
\author{Travis C. Briles}
\author{H. J. Lewandowski}
 \email{lewandoh@colorado.edu}
 \affiliation{Department of Physics, University of Colorado, Boulder, CO 80309, USA}
  \affiliation{JILA, National Institute of Standards and Technology and University of Colorado, Boulder, CO, 80309, USA}

\date{\today}%

\begin{abstract}
Vacuum ultraviolet (VUV) light at 118 nm has been shown to be a powerful tool to ionize molecules for various gas-phase chemical studies. A convenient table top source of 118 nm light can be produced by frequency tripling 355 nm light from a Nd:YAG laser in xenon gas. This process has a low efficiency, typically producing only nJ/pulse of VUV light. Simple models of the tripling process predict the power of 118 nm light produced should increase quadratically with increasing xenon pressure.  However, experimental 118 nm production has been observed to reach a maximum and then decrease to zero with increasing xenon pressure. Here, we describe the basic theory and experimental setup for producing 118 nm light and a new proposed model for the mechanism limiting the production based on pressure broadened absorption.
\end{abstract}

                           
\maketitle

\section{\label{sec:Introduction} Introduction}

Vacuum ultraviolet (VUV: $\lambda \approx 100$-$200$~nm) light sources \cite{samson1967techniques, samson1998vacuumV1,samson1998vacuumV2} have important applications in a wide variety of experiments in atomic physics and physical chemistry, including molecular spectroscopy for the understanding of chemical bonds, \cite{Ng2002, albert2013studies} high resolution studies for metrology, \cite{cingoz2012, de2011high, hannemann2006frequency, rellergert2010constraining, eyler2008prospects} and as a universal ionization source in mass spectrometry\cite{pallix1989,steenvoorden1991}. High fluxes of VUV light are available from synchrotrons, but the complexity, cost, and beam-time constraints of these sources makes a more accessible tabletop VUV source desirable.  With the exception of excimer lasers,\cite{basting2005excimer} which operate at only a few set wavelengths, tabletop sources of VUV light require use of nonlinear optical techniques.  Nonlinear optical frequency mixing, \cite{BoydNonLinearBook, SHENnonlinearBOOK, new2011introduction} of which harmonic generation is a special case, is a powerful technique for producing coherent radiation at wavelengths where small commercial laser sources do not typically operate.\cite{Couch20}

One such VUV source is made by frequency tripling the third harmonic output of a Nd:YAG laser in a xenon-argon gas mixture to create 118 nm light. The 118 nm photon can be used in single or multi-photon ionization applications. Specifically, chemical dynamics studies use 118 nm light in ion-pair photodissociation spectroscopy,\cite{Suto1997} ion pair production,\cite{Munakata1989} anion photoelectron spectroscopy,\cite{Yang2008} and one-photon ionization processes. \cite{Li1999,Jackson2000,Shi2002,Shin2004} Additionally, 118 nm photons have been used to perform multi-photon ionization of cold radical molecules.\cite{Beames2011,Beames2014,Gray2017} 

Since the conversion efficiency to VUV light is low ($\sim 10^{-5}$)\cite{McCann1988} and may constrain the detection sensitivity of an experiment, it is important to understand the limits of the tripling technique in order to experimentally optimize 118 nm light production. The basic theory of 118 nm generation in a xenon-argon gas cell \cite{ward1969optical,MILESharris1973,bjorklund1975effects} and experimental setup and challenges\cite{Lockyer1997} have been detailed in previous papers. However, the experimentally observed 118 nm photon production has been observed to significantly diverge from theory at high gas pressures and has not been modeled so far.

Here, we describe the basic theory and experimental setup for generating 118 nm light and characterize the limitation of the VUV production at high xenon pressures. We propose absorption at 118 nm by xenon as the mechanism that limits 118 nm production. We develop a model that accounts for this absorption, whose predictions are in good agreement with our experimental results. 

\section{\label{sec:nonresonantTHG} Theory of Non-Resonant Third Harmonic Generation}

Anisotropic crystals with large second-order nonlinear optical susceptibilities ($\chi^{(2)}$) and a wide optical transparency range can be efficient at producing visible, near-infrared, and soft UV light. Since these crystals start to absorb light below $200$~nm, generation of VUV light via nonlinear optical frequency mixing generally requires a gaseous nonlinear medium, which is transparent at these wavelengths. However, the structural symmetry of isotropic media such as gases means that only odd-order nonlinear susceptibilities are non-vanishing,\cite{KLEINMAN1962} leaving the third-order susceptibility, $\chi^{(3)}$, as the first nonlinear term. \cite{BoydNonLinearBook}  

Third-harmonic light produced by the large $\chi^{(3)}$ in these systems can extend into the VUV and extreme ultraviolet regions of the spectrum. For this process to be efficient, phase matching is important, as the fundamental light and third-harmonic light must maintain a fixed phase relationship in order to coherently build the tripled light. In the gaseous system discussed here, phase matching is accomplished by mixing a negatively dispersive gas that acts as the nonlinear medium with positively dispersive gas that can be used to tune the phase matching. In the remainder of this section, we discuss the basic model of third-harmonic generation (THG) and conditions to optimize phase matching.

\begin{figure}[htbp]
\centering
\includegraphics[width=3.5in]{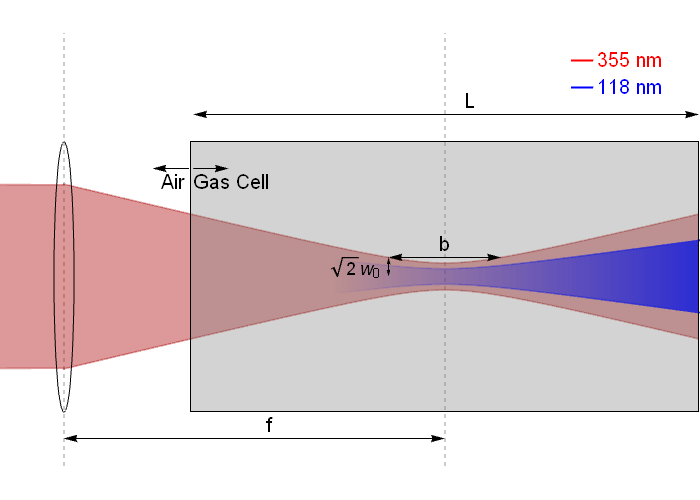}
\caption{\label{fig:THGdiagram}Diagram showing the parameters of the gas cell for generating 118 nm light (blue). The 355 nm beam (red) is focused into the gas cell (gray box) by a lens with focal length $f$. The confocal parameter $b$ and beam waist $w_0$ are noted.
}
\end{figure}

The basic setup for THG is depicted in Fig. \ref{fig:THGdiagram}. The fundamental light at 355 nm is focused into a gas cell of length, $L$, with a beam waist of $w_0$ and a confocal parameter of $b=\frac{2\pi w_0^2}{\lambda}$.  The third-harmonic light is shown in blue. The total power of the third-harmonic light produced by a focused Gaussian beam with incident fundamental power $\mathcal{P}_\omega$ is given by \cite{MILESharris1973,bjorklund1975effects} 

\begin{equation}
\mathcal{P}_{3\omega}  = \frac{8.215 \times 10^{-2}}{\left(3\lambda \right)^4} \left\vert \chi_{eff}^{(3)} \right\vert^2 \mathcal{P}^3_{\omega} \left\vert F_1 (b \Delta k) \right\vert^2.
\label{Eq:GENERALfourWAVEmixingPOWERd}
\end{equation}

\noindent Here, all powers are in watts, $\lambda$ is the third-harmonic wavelength in $nm$, and $\chi_{eff}^{(3)}$ is the macroscopic nonlinear coefficient. $F_1 (b \Delta k)$ is a dimensionless function that describes phase matching, and $\Delta k$ is the wave-vector mismatch between the fundamental and third-harmonic light. $\Delta k$ is the term responsible for phase matching and optimizing it improves the efficiency of the THG process. 

Experimentally, there are three parameters that can be varied to optimize the third-harmonic power (Eq. \ref{Eq:GENERALfourWAVEmixingPOWERd}): (1) maximizing the macroscopic nonlinear coefficient $\chi_{eff}^{(3)}$, (2) maximizing $ F_1(b\Delta k)$ by optimizing phase matching, and (3) maximizing the power in the fundamental ($\mathcal{P}_{\omega}$). However, as shown below, independent maximization of the first two parameters in a single gas is impossible because both $\chi^{(3)}_{eff}$ and $F_1(b\Delta k)$ are dependent on $\mathcal{N}$, the atomic number density of the gas. $\chi_{eff}^{(3)}$ is given by
\begin{equation}
\chi_{eff}^{(3)} = \mathcal{N}\times \chi_a^{(3)},
\end{equation}

\noindent where $\mathcal{N}$ is in units of atoms/cm$^3$ and $\chi_a^{(3)}$ is the third-order nonlinear coefficient per atom in $esu$. Generally $F_{1}(b\Delta k)$  must be evaluated numerically. However, in the tight-focusing limit, where the nonlinear medium is much longer than the confocal parameter of the focused fundamental beam ($b \ll L$), $F_{1}(b\Delta k)$ can be solved analytically to give \cite{bjorklund1975effects, ward1969optical} 
 
\begin{equation}
\left\vert F_{1}(b\Delta k)\right\vert^2 = \begin{cases}
\pi^2 (b \Delta k)^2 e^{b \Delta k}& \text{$\Delta k < 0,$} \\
0 &\text{$\Delta k \geq 0,$}
\end{cases}
\label{Eq:GENERALthgPHASEmatchingFUNCTION}
\end{equation}

\noindent which has a maximum at $\Delta k_{max} = -\frac{2}{b}$. 

$\Delta k$ is related to $\mathcal{N}$ through the index of refraction, and is given by

\begin{eqnarray}
\Delta k &=& \frac{2\pi}{\lambda} \left[n_1 - n_3 \right].
\label{Eq:THGwaveVECTORmismatch}
\end{eqnarray}

\noindent In this notation, $n_1$ and $n_3$ are the index of refraction at the third-harmonic wavelength $\lambda$ and the fundamental wavelength $3\lambda$, respectively. For a gas, the effective macroscopic index of refraction, $n$, is given by 

\begin{equation}
n(\lambda) = 1 + \mathcal{N} \chi_{a}^{(1)} (\lambda),
\label{Eq:SELLMEIERleadUP}
\end{equation}

\noindent where $\chi_a^{(1)}$ is the atomic linear susceptibility at wavelength $\lambda$. Therefore, $\Delta k$ can be expressed as

\begin{equation}
\Delta k = C \mathcal{N}, 
\label{Eq:GENERICdefnWAVEVECTORmismatchPERatomTHG}
\end{equation}

\noindent where $C =  \frac{2\pi}{\lambda} \left[  \chi^{(1)}_a (\lambda) -  \chi^{(1)}_a (3\lambda) \right]$ is a constant describing the microscopic dispersive properties of the gaseous medium.  $C$ can be thought of as the effective wave-vector mismatch per atom. Phase matching is adjusted by tuning $\Delta k$, and thus $F_1(b\Delta k)$, via the pressure of the gas.

As can be seen from Eqn. \ref{Eq:GENERALthgPHASEmatchingFUNCTION}, $\Delta k$ must be negative to generate third-harmonic light. This occurs only in media that exhibits negative dispersion (i.e.,  $n_3 > n_1$). Xenon was chosen as the nonlinear medium for THG of 118 nm because it is the only rare gas that exhibits negative dispersion at this wavelength. With only a negatively dispersive gas, $\Delta k$ and $\chi^{(3)}_{eff}$ are both dependent on $\mathcal{N}$. If a positively dispersive gas with negligible nonlinearity is added, $\Delta k$ and $\chi^{(3)}_{eff}$ can be decoupled. Argon is typically used as the positively dispersive gas for THG of 118 nm light.\cite{Lockyer1997} For a mixture of the two gases, the total wave-vector mismatch at the nominal third-harmonic wavelength, $\lambda_0$, is given by summing over the partial pressures of each gas, 

\begin{eqnarray}
\Delta k (\lambda_0) &=&  \mathcal{N}_{Xe} C_{Xe}(\lambda_0) + \mathcal{N}_{Ar} C_{Ar} (\lambda_0) \nonumber,
\\
&=&  \mathcal{N}_{Xe} \left[C_{Xe}(\lambda_0) + R C_{Ar} (\lambda_0) \right],
\label{Eq:wavevectorMISMATCHtwoCOMPONENTgas1}
\end{eqnarray}

\noindent where $R=\frac{\mathcal{N}_{Ar}}{\mathcal{N}_{Xe}}$ is the pressure ratio of the two gases. With the addition of argon, $\chi^{(3)}_{eff}$ can be adjusted using $\mathcal{N}_{Xe}$, while $\Delta k$ can be maximized using $\mathcal{N}_{Ar}$. 

Since $F_1(b\Delta k)$ is maximized when $\Delta k (\lambda_0)_{max} = -\frac{2}{b}$, phase matching is optimized when the pressure ratio is 

\begin{equation}
R_{\text{opt}} = \frac{-2}{b} \frac{1}{\mathcal{N}_{Xe}  C_{Ar} (\lambda_0)  } - \frac{C_{Xe} (\lambda_0)}{C_{Ar} (\lambda_0)}.
\label{Eq:pressureratio}
\end{equation}

\noindent For large xenon pressures, which are desirable for increasing $\chi^{(3)}_{\text{eff}}$ and thus $\mathcal{P}_{3\omega}$, the optimum ratio approaches the constant value of $\frac{C_{Xe}}{C_{Ar}}$ and is independent of the value of confocal parameter, $b$. Using the values $C_{Xe}(118.2~$nm$) = -6.12 \times 10^{-17} $cm$^2$ and $C_{Ar}(118.2~$nm$) = +5.33 \times 10^{-18} $cm$^2$ from Mahon \textit{et al.,} \cite{MAHON1979} this ratio is 11.48.

\section{Experimental Setup for Third-Harmonic Generation}
 
\begin{figure}[htbp]
\centering
\includegraphics[width=3.5in]{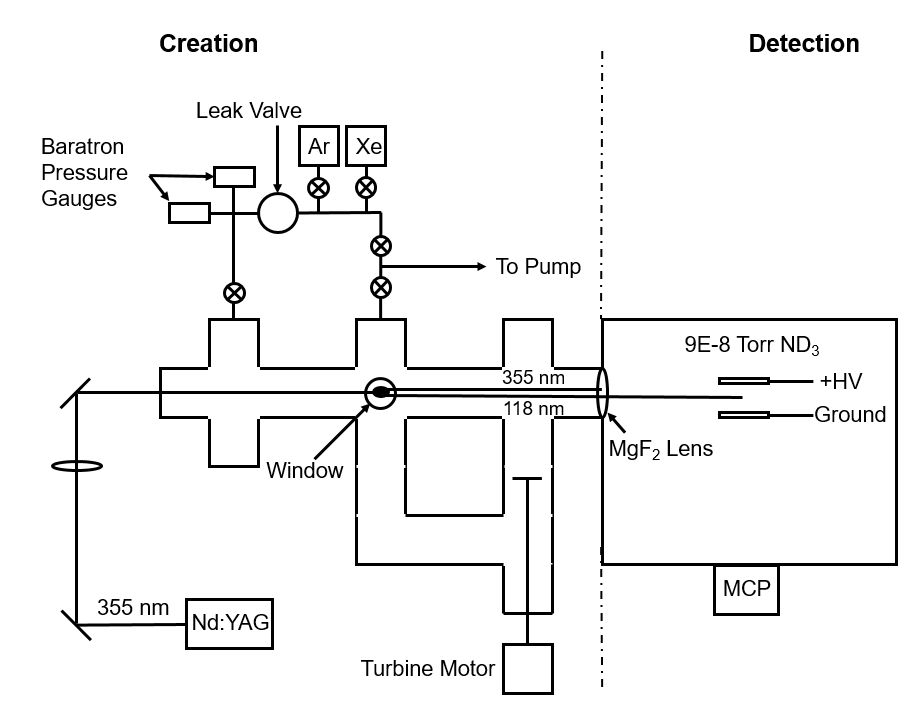}
\caption{\label{fig:ExperimentalSetup}Schematic of the experimental setup. A lens focuses the 355 nm light into the gas cell where 118 nm light is created. A window placed above the focus of the 355 nm beam allows perpendicular optical access to the tripling region. The MgF$_2$ lens focuses the 118 nm light and defocuses the 355 nm in the detection region. 118 nm flux is detected via ionization of deuterated ammonia (ND$_3$). A low flow rate of ND$_3$ fills the detection chamber to a partial pressure of $9\times 10^{-8}$ Torr. ND$_3^+$ ions are accelerated through a time-of-flight mass spectrometer and detected by a micro-channel plate (MCP). 
}
\end{figure}

The experimental apparatus used to generate VUV light is shown in Fig. \ref{fig:ExperimentalSetup}. Light at 355 nm  from a Q-switched Nd:YAG laser (Continuum: Inlite III-10, internally tripled) is focused into a mixture of xenon and argon contained in a stainless steel cell using a 50 cm focal length lens. The gas cell is made of standard 304-SS conflat vacuum parts with an overall length of $\sim40$ cm. A UV fused silica window seals the input end of the cell, where the 355 nm light enters. The output end, between the gas cell and the UHV detection chamber, is sealed using a 25 cm focal length MgF$_2$ lens, which also serves to focus the VUV light into the detection region.

The focal length of the lens and size of the cell must be chosen to avoid burning optics and maximize gas conductance. The cell is three standard conflat crosses connected together and has a length of $\sim$ 40 cm. The focal length of the 50 cm lens, external to the cell, was chosen to be approximately the length of the gas cell. The focus is designed to be in the center of the gas cell, so that the 355 nm intensity at the input window and output lens are low enough to prevent damaging the optics. The position of the focus also allows for perpendicular optical access through a conflat window into the focal region. Although LiF windows have a higher transmission at 118~nm, a MgF$_2$ window was chosen for the output lens due to its hardness and because it is not hygroscopic, allowing the window to handle pressure differentials in excess of 1 atm. The large positive dispersion in MgF$_2$ causes the focal length of the lens at 118 nm to be about $58\%$ shorter than that at 355 nm.  This property is exploited to minimize the intensity of 355 nm in the detection region by defocusing the 355 nm light, which helps to prevent multi-photon processes. Additionally, most of the defocused 355 nm light is blocked from even reaching the detection region by a macor iris placed after the MgF$_2$ lens. 

The gas cell is constructed from 2-3/4 inch conflat pieces to provide adequate gas conductance. Attempts at making portions of the gas cell out of 1/4 inch ID tubing led to trapped pockets of unmixed xenon, which slowly diffused out over several days. The release of trapped xenon led to drifts in the gas ratio, changing the phase matching and leading to drifts in the amount of 118 nm light produced. 

Xenon and argon are introduced into the gas cell through a leak valve and needle valve. The needle valve is important for separating the gas mixing cell from the low conductance elements such as the pressure gauges and gas lines connecting the leak valve. Leaving open the needle valve between the gas mixing cell and leak valve increases the gas mixing time by a factor of three.

Gas mixing in the cell can be aided by a turbine, which causes gas to flow in a loop of 2-3/4 inch conflat pieces (Fig. \ref{fig:ExperimentalSetup}).  Gas mixing using the turbine did not reach equilibrium noticeably faster than without the turbine, but the 118 nm flux varied more smoothly while mixing. Without the turbine, adding argon to the system would cause large spikes and dips in the measured 118 flux, as the introduction of argon rapidly changed the Xe/Ar ratio in the tripling region. Unaided mixing in the cell took only 15 minutes with no signs of continued mixing or signal loss observed over the course of several weeks. 

 The 118 nm light is detected indirectly by using it to ionize deuterated ammonia (ND$_3$) and subsequently detecting ND$_3^+$. 118 nm light is focused by the MgF$_2$ lens between two electrodes in the UHV chamber. Ionized ND$_3$ is accelerated through a time-of-flight mass spectrometer (TOF-MS) and is detected by a micro-channel plate (MCP) (Fig. \ref{fig:ExperimentalSetup}). The number of ND$_3^+$ ions produced is proportional to the number of 118 nm photons that are focused into the detection region, since the ionization is a one-photon process and no saturation effects are observed. 

\section{Phase Matching and Conversion Efficiency Limitations}

\begin{figure}
\centering
\includegraphics[width=3.5in]{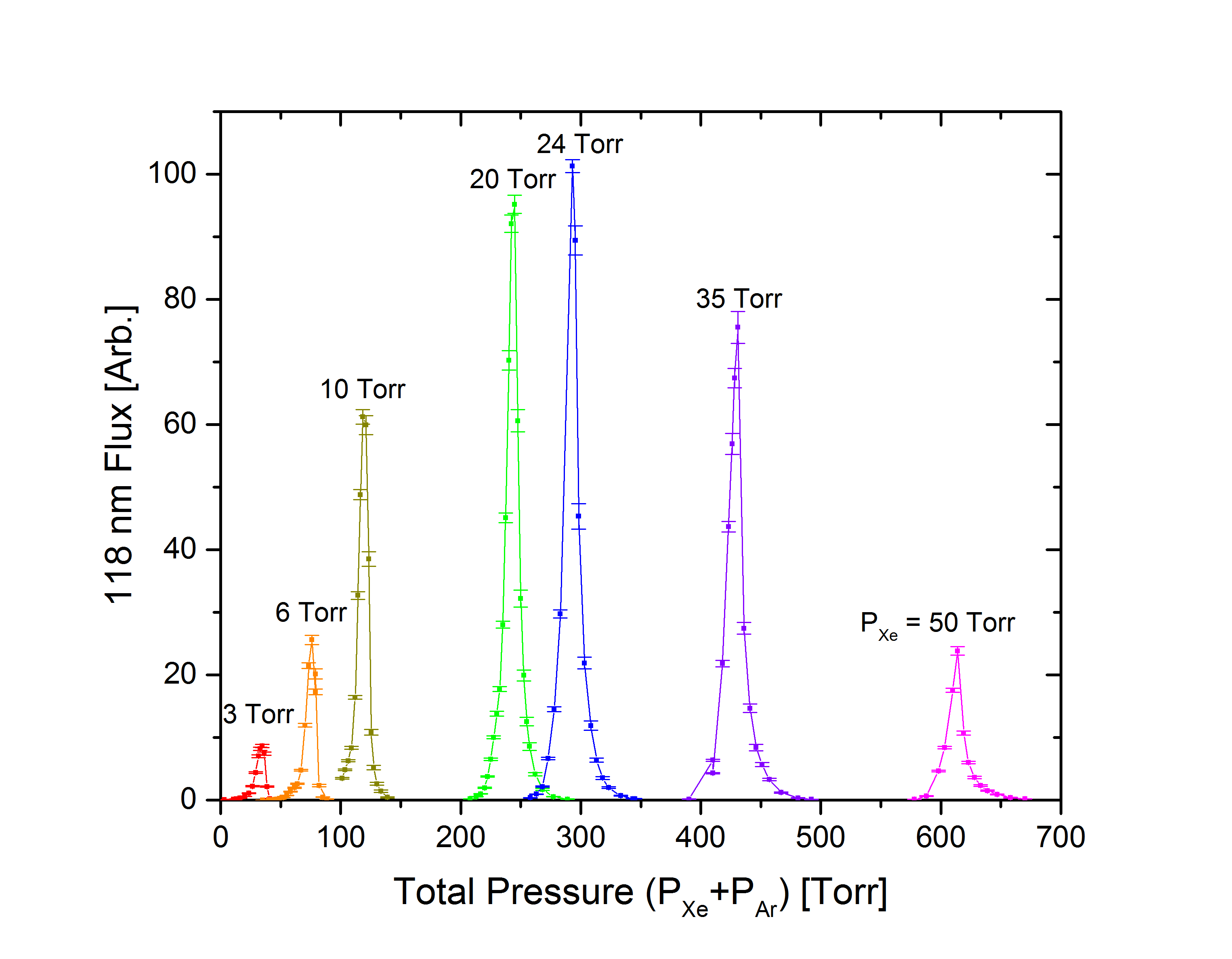}
\caption{\label{fig:118nmPHASEmatching}Phase-matching curves for different pressures of xenon (labeled above). 118nm flux is measured while Ar is added to a fixed pressure of Xe and plotted as a function of total pressure.
}
\end{figure}

\begin{figure}[htbp]
\centering
\includegraphics[width=3.5in]{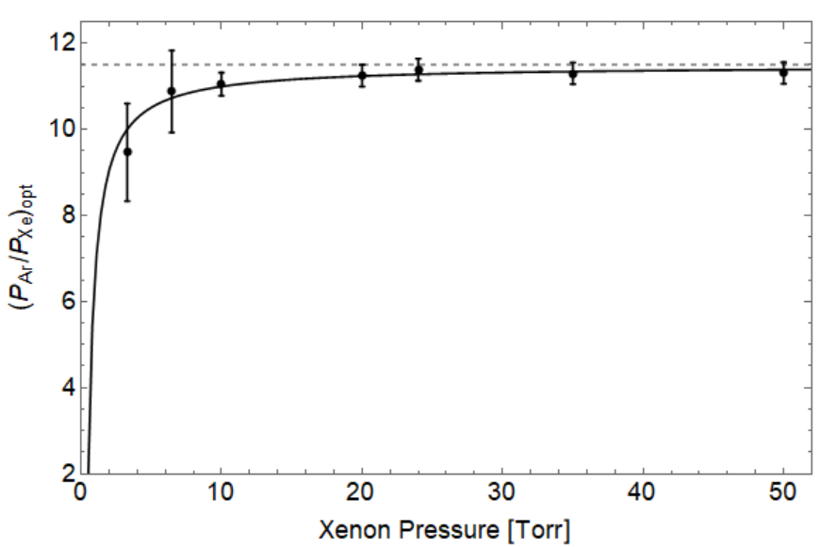}
\caption{\label{fig:PressureRatio}  Plot of optimum Ar:Xe ratio vs. Xe pressure. The solid line is the output of a fit of Eq. \ref{Eq:pressureratio} to the data with b $= 2.3 \pm 0.3$ cm as the only fit parameter. The dashed line represents the asymptotic limit of 11.48 predicted from the model. \
}
\end{figure}

To experimentally determine the phase matched ratio of partial pressures of Xe/Ar, we add a set amount of xenon to the mixing cell and then add increments of argon in order to map out each phase-matching curve.  Fig. \ref{fig:118nmPHASEmatching} shows phase-matching curves for several xenon pressures. Each phase-matching curve was fit to Eq. \ref{Eq:GENERALthgPHASEmatchingFUNCTION} to extract the location of the peak 118 nm flux and the optimum Xe/Ar ratio for a given xenon pressure. The optimum Xe/Ar ratios are plotted in Fig. \ref{fig:PressureRatio} along with a fit to Eq. \ref{Eq:pressureratio}, which gives $b= 2.3 \pm 0.3$ cm. As shown in Eq. \ref{Eq:pressureratio}, the optimum Xe/Ar ratio depends on the confocal parameter and xenon pressure, but for high xenon pressures the ratio approaches the asymptotic limit of 11.48 (horizontal dashed line). 

Although the optimum pressure ratios are consistent with the model, the conversion efficiency of 355 nm light to 118 nm does not agree with model predictions. Eq. \ref{Eq:GENERALfourWAVEmixingPOWERd} predicts that for a phase-matched mixture, the 118 nm light produced should scale as the square of the xenon pressure. However, mixtures with more than 24 Torr of xenon (290 Torr total pressure) yielded decreasing 118 nm light with increasing Xe pressure. At xenon pressures $>$70 Torr, almost no 118 nm light was detected. This same turnover in 118 nm light production versus total pressure has been observed in Shi \textit{et al.}. \cite{Shi2002} In other studies, optimal conversion efficiencies have been observed at lower total pressures of 20 Torr,\cite{Jackson2000} 50 Torr,\cite{Yang2008} and 200 Torr. \cite{Shin2004} The prevalence of optimal conversion efficiencies at low total pressures and the large discrepancy between the predicted and measured yield of 118 nm light at high pressures indicates that some process, which is not included in the models, is significantly limiting the production of 118 nm light.

\begin{figure}[htbp]
\centering
\includegraphics[width=3.5in]{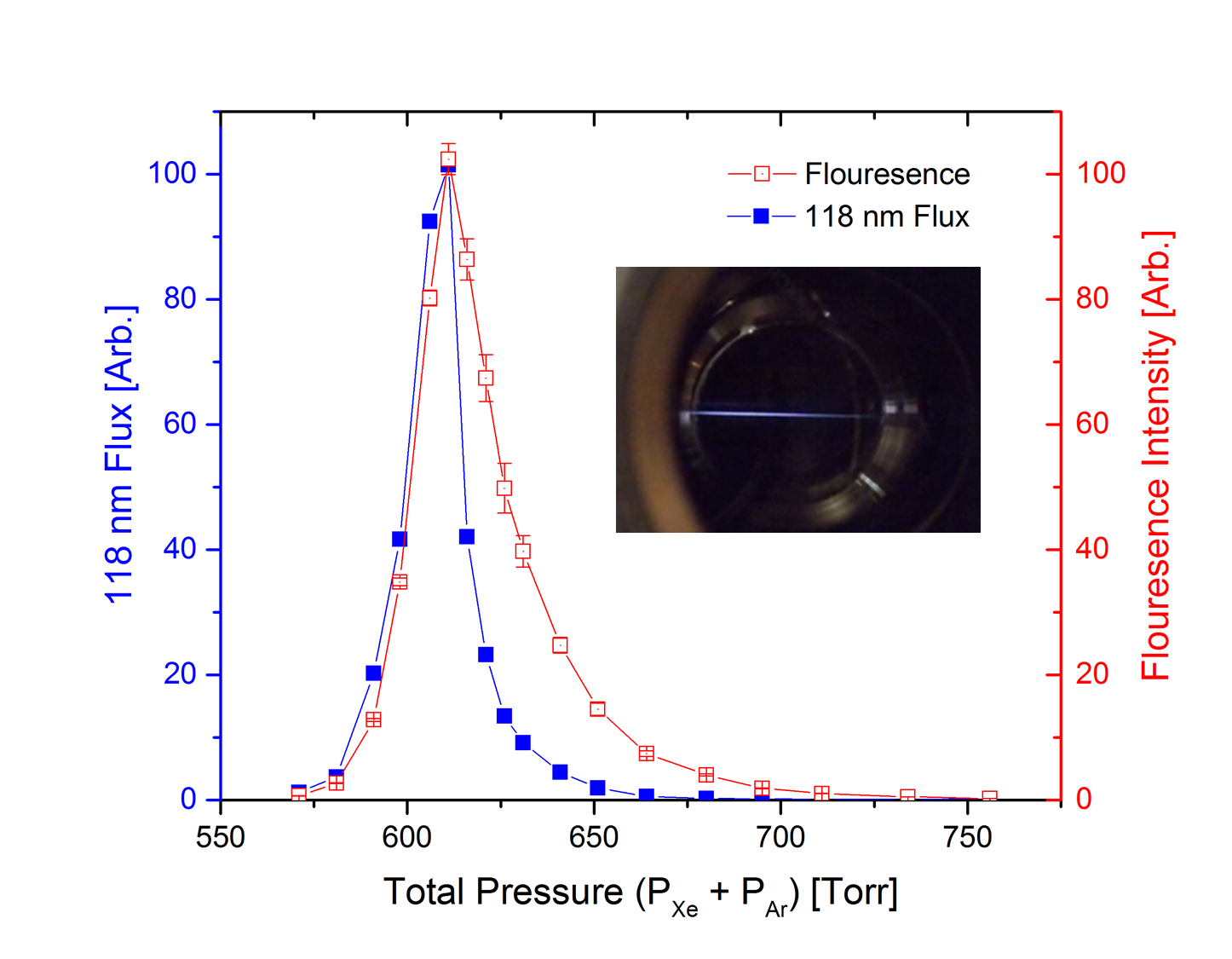}
\caption{\label{fig:FluorescencePhaseMatch} Plot of 118 nm flux and the fluorescence signal vs total pressure for $P_{Xe} = 50$ Torr. The fluorescence signal is correlated with the presence of 118 nm light. The inset show a picture of the fluorescence streak through a conflat window.}
\end{figure}

\begin{figure}[htbp]
\centering
\includegraphics[width=3.5in]{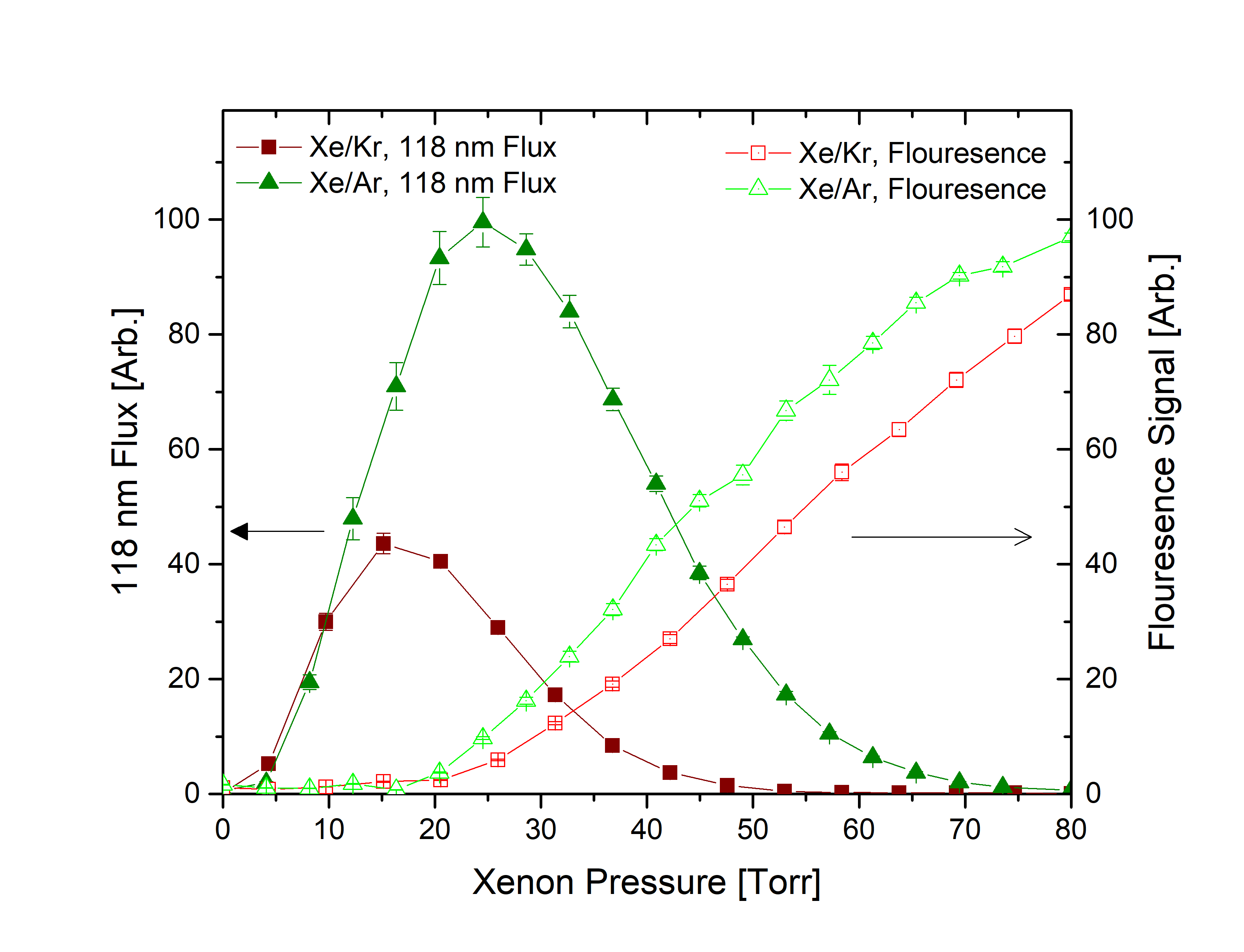}
\caption{\label{fig:Turnover} Comparison of the 118 nm signal and fluorescence streak signal at different xenon pressures for a phase-matched mixture with Ar (Ar:Xe Ratio = 11.3) and Kr (Kr:Xe Ratio = 4.6).  
}
\end{figure}

The deviation from theory for the expected 118 nm flux of a phase-matched mixture of Xe/Ar at higher total pressures is accompanied by the appearance of a fluorescence streak inside the gas cell (inset Fig. \ref{fig:FluorescencePhaseMatch}). It appears near the focus of the 355 nm light and along the propagation axis. The fluorescence streak does not appear when 355 nm light is focused in to the cell filled with only argon or only xenon. Instead it appears only when xenon and argon are both present and near a phase-matched ratio. The fluorescence is collected through a vaccuum window using a lens outside the gas cell to focus the light onto a photodiode. The fluorescence intensity follows the shape of the generated VUV light flux throughout a phase-matching curve as shown in Fig. \ref{fig:FluorescencePhaseMatch}, indicating that 118 nm photons play a role in the fluorescence. Fig. \ref{fig:Turnover} shows the 118 nm flux and fluorescence signal as a function of the xenon pressure for a phase-matched mixture. The fluorescence appears to begin at some threshold xenon pressure and increases roughly linearly in intensity with increasing xenon pressure. 

\begin{figure}[htp]

\includegraphics[width=3.5in]{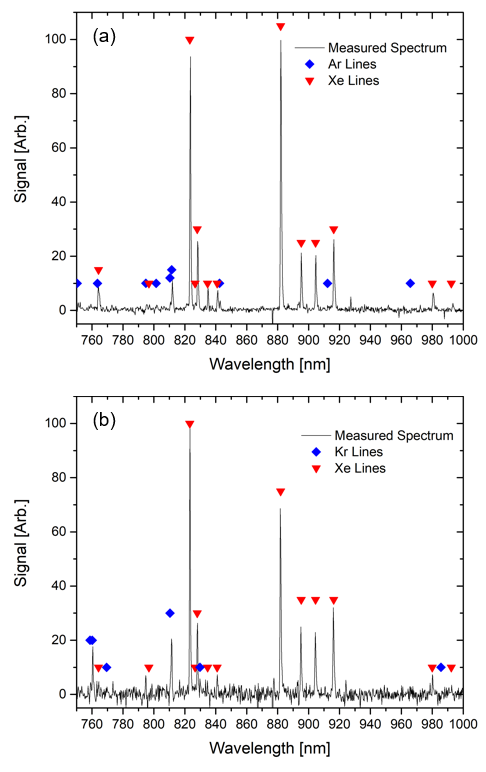}

\caption{\label{fig:ArKrSpectra}Spectra of the fluorescence streak created in a phase-matched (a) Xe:Ar mixture (Ar:Xe Ratio = 11.3) and (b) Xe:Kr mixture (Kr:Xe Ratio = 4.6). Lines corresponding to Ar I, Kr I, and Xe I are assigned according to criteria described in the text.}

\end{figure}

118 nm light was also created using krypton as the phase-matching gas. Using the same apparatus and measurements described above, xenon and krypton were phase matched with an optimum Kr/Xe ratio of 4.6, which agreed with the krypton dispersion calculated in Mahon \textit{et al.} \cite{MAHON1979} This mixture also produced a maximum of 118 nm light followed by a decrease with increasing gas pressure. The fluorescence streak was also present only with a phase-matched mixture of the two gasses. (Fig. \ref{fig:Turnover}). 
Spectra of the fluorescence were obtained for the Xe/Ar (Fig.\ref{fig:ArKrSpectra} (a)) and Xe/Kr (Fig.\ref{fig:ArKrSpectra} (b)) systems using an Ocean Optics USB2000 spectrometer. Spectral lines were assigned using data from the NIST Atomic Spectra Database. \cite{NIST_ASD} We mark the location of all lines in the NIST database that are above a certain threshold relative intensity \cite{NIST_ASD}  in order to identify observed lines and note which lines are missing from the spectra. For Xe and Kr, a threshold of 500 was used, and for Ar a threshold of 20000 relative intensity was used. The Xe/Ar spectrum (Fig. \ref{fig:ArKrSpectra} (a)) shows predominantly Xe I spectral lines and weak Ar I spectral lines. The lines are similar in relative signal to those observed from a xenon arc lamp, indicating that the fluorescence is likely from electron-ion recombination in xenon. The spectrum for the Xe/Kr system (Fig.\ref{fig:ArKrSpectra} (b)) also shows predominantly Xe I spectral lines, with Kr I spectral lines appearing relatively weakly. Xenon is ionized preferentially despite being present in a lower concentration than the phase-matching gasses (Ar or Kr). This would seem to indicate that ionization is not due to just the high electric field of the 355 nm light, but a resonant process involving xenon.
The energetics suggest a two-photon process is causing the ionization of xenon. The energy of a single 118 nm photon (10.5eV) is below the ionization potential of xenon (12.1 eV). However, the combined energy of one 118 nm photon and one 355 nm photon (14 eV) is enough to ionize xenon, while being below the ionization potential of argon (15.8 eV). The ionization potential of krypton (14 eV) is low enough to possibly be ionized in this system, however strong ionization of kryton is not evident. Two-photon ionization of xenon, where one of the photons is 118 nm and the other is a 355 nm photon, is likely the mechanism producing xenon ions. The intensity of the 355 nm light is much larger than that of the third harmonic, so we can assume the absorption of the 118 nm light is the rate limiting step. We model this process as two-photon, two-color ionization model where the absorption of the 118 nm photon is far from resonance and is affected by the pressure in the gas cell due to pressure broadening. This pressure broadened absorption is the proposed mechanism for loss of 118 nm light in the detection region for high cell pressures. 
In Xenon the two lines closest to resonance with the 118 nm light are the Xe ($^2P_{3/2}$)5d $\left[\frac{3}{2}\right]$ line at 119.2 nm and the Xe ($^2P_{3/2} $)7s line at 117 nm. We also must consider the Kr ($^2P_{1/2}$)5d $\left[\frac{1}{2}\right]$ resonance line at 116.5 nm, which can lead to additional absorption in the case of phase matching with Kr. Argon does not have any transitions near enough to 118 nm to cause significant loss.

\section{\label{sec:CrossSec} Absorption Cross Section}
We consider far off-resonance absorption in the presences of pressure broadening.  The atomic absorption cross section for light at frequency $\omega$ due to an absorption line at $\omega_0$ is given by 
\begin{equation}
    \sigma(\omega) = \frac{f e^2}{m \epsilon_0 c \gamma} \frac{(\gamma/2)^2}{(\omega-\omega_0)^2 + (\gamma/2)^2},
\label{Eq:crosssection}
\end{equation}

\noindent where $f$ is the oscillator strength, $e$ is the electron charge, $m$ is the mass of the electron, $\epsilon_0$ is the permittivity of free space, $c$ is the speed of light, and $\gamma$ is the linewidth. Far from resonance, $(\omega-\omega_0)^2 >> (\gamma/2)^2$, and Eqn. (\ref{Eq:crosssection}) can be approximated as 

\begin{equation}
\sigma(\omega) = \frac{f e^2}{4 m \epsilon_0 c} \frac{\gamma}{(\omega-\omega_0)^2},
\label{Eq:crosssection2}
\end{equation}

\noindent noting that the cross section is proportional to the linewidth. 
For gasses at high pressures, collisions between atoms contribute to the Lorentzian linewidth of an absorption in a process known as pressure broadening. Neutral atoms undergoing collisions induce a dipole that perturbs the energy levels and broadens the linewidth of transitions. Collisions between atoms of the same species experience resonant collisions that enhance this broadening. Same-species broadening is typically $\sim 1 $ GHz$ \times 2\pi$ per Torr of gas vs. $\sim 10 $ MHz$ \times 2\pi$ per Torr of gas for collisions between different species.\cite{Hebner1991,Pitz2009}  In our two-species gas, although the phase matching gas is present in much higher quantities than xenon, resonant Xe-Xe collisional broadening will dominate the pressure broadening. The pressure broadened linewidth is given by 
\begin{equation}
 \gamma = \gamma_n + \gamma_c,
\label{Eq:gammas}
\end{equation}

\noindent where $\gamma_n$ is the natural linewidth and $\gamma_c$ is the collisional linewidth, which is proportional to the density, $\gamma_c = \beta \mathcal{N}$, where $\beta$ is the collisional broadening coefficient and $\mathcal{N}$ is the density of the gas. For the lines considered in this paper $\gamma_n \sim 100 $ MHz$ \times 2\pi$, while $\gamma_c \sim 1 $ GHz$ \times 2\pi$ per Torr of gas.\cite{Ferrell1987} So, for pressures used in this experiment $\gamma_n<<\gamma_c$, and $\gamma_n$ can be ignored such that $\gamma \cong \gamma_c$. This allows us to write our pressure broadened cross section as 

\begin{eqnarray}
   \sigma(\omega) &=& \frac{f e^2}{4 m \epsilon_0 c} \frac{\beta \mathcal{N}}{(\omega-\omega_0)^2}\nonumber,
   \\
   &=& \alpha \mathcal{N},
\label{Eq:crosssectionalpha}
\end{eqnarray}

\noindent where $\alpha$ is defined as our pressure broadening constant.

\section{\label{sec:Simulation} Simulation Results}
In order to compare our pressure broadened absorption model to the experimental data, we simulate the simultaneous tripling and absorption processes in our gas system. The electric field of an electromagnetic wave with frequency, $\omega$, and wavenumber, k, propagating along the z-axis is given by 

\begin{equation}
E(z,t)=A e^{i(k z - \omega t)} + c.c.,
\label{Eq:ElectricField}
\end{equation}

\noindent where A is the electric field amplitude. The intensity of the electric field is related to the electric field amplitude by $I = 2 n \epsilon_0 c |A|^2$. The change in the intensity of the 118 nm beam as it propagates through a medium can be simulated by calculating the change in the electric field amplitude.  The differential equation describing the propagation of the 118 nm beam is given by 

\begin{equation}
\frac{d|A_{118}|}{dz} =\frac{1}{2} \chi^{(3)}\mathcal{N}_{Xe}k_{118}|A_{355}|^3 - \frac{1}{2} \sigma \mathcal{N}_{Xe} |A_{118}|,
\label{Eq:Model1}
\end{equation}

\noindent where $\chi^{(3)}$ is the third order susceptibility, $\mathcal{N}_{Xe}$ is the density of xenon, $k_{118}$ is the wavenumber of the 118 nm light, $A_{355}$ is the field amplitude of the 355 nm beam, $\sigma$ is the absorption cross section, and $A_{118}$ is the field amplitude of the 118 nm beam.  The first term describes the third harmonic generation of 118 nm light assuming perfect phase matching and the second term describes the loss of 118 nm light due to absorption. Using Eqn (\ref{Eq:crosssectionalpha}), this becomes 

\begin{equation}
\frac{d|A_{118}|}{dz} = \frac{1}{2} \chi^{(3)}\mathcal{N}_{Xe}k_{118}|A_{355}|^3 - \frac{1}{2} \alpha \mathcal{N}_{Xe}^2 |A_{118}|.
\label{Eq:Model2}
\end{equation}
 
The 355 nm power is kept constant, assuming the third harmonic generation leads to negligible loss. To account for the fact that the 355 nm beam is focusing through the gas cell, the electric field amplitude along the z axis is scaled by the radius of the beam
\begin{equation}
|A_{355}(z)| = |A_{355}(0)| \frac{\omega_0}{\omega(z)},
\label{Eq:355Scaling}
\end{equation} 
where $\omega_0$ is the beam waist, $\omega(z)=\omega_0\sqrt{1+(\lambda z / \pi \omega_0^2)^2}$ is the beam radius as a function of z, and $A_{355}(0)$ is the field amplitude at the focus calculated from experimental parameters. 

Although the 118 nm beam also has a transverse spread that changes through the focus, the 118 nm amplitude is not scaled by the beam radius. This assumes that the effects of the transverse spread are negligible on relative measurements across our parameter space. The majority of the third harmonic generation takes place within the Rayleigh range, where the beam radius varies from $\omega_0 - \sqrt{2}\omega_0$. We assume this variation is negligible and the 118 nm light is created with roughly the same beam radius. Linear absorption does not affect the width of a Gaussian beam, thus the transverse expansion of the beam is the same for all parameters, and ignoring the expansion does not effect relative measurements.

The differential equation is numerically integrated over the 40 cm length of the gas cell, with the focus at the center. The 118 nm field amplitude at the end is used to calculate a beam flux, since our experimental results are a measure of the flux of the 118 nm photons. Additionally, we assume that the amount of light from the fluorescence streak is proportional to the loss of 118 nm light in a region around the focus of the 355 nm beam that approximates our fluorescence collection region.


 Using our simulations, we want to model the experiment shown in Fig (\ref{fig:Turnover}), then compare the simulated and experimental results to extract a value for $\alpha$. The two free parameters in this simulation are the pressure-broadening constant, $\alpha$, and an overall vertical scaling of the magnitude of the signal. For a single value of $\alpha$, the simulation is run for a range of xenon pressures to produce a curve of 118 nm flux vs pressure similar to Fig (\ref{fig:Turnover}). The simulated curves are fit to the experimental data by calculating a sum of squared residuals between the experimental and simulated results for a given value of $\alpha$ and vertical scaling factor. 

\begin{figure}[htbp]
\centering
\includegraphics[width=3.5in]{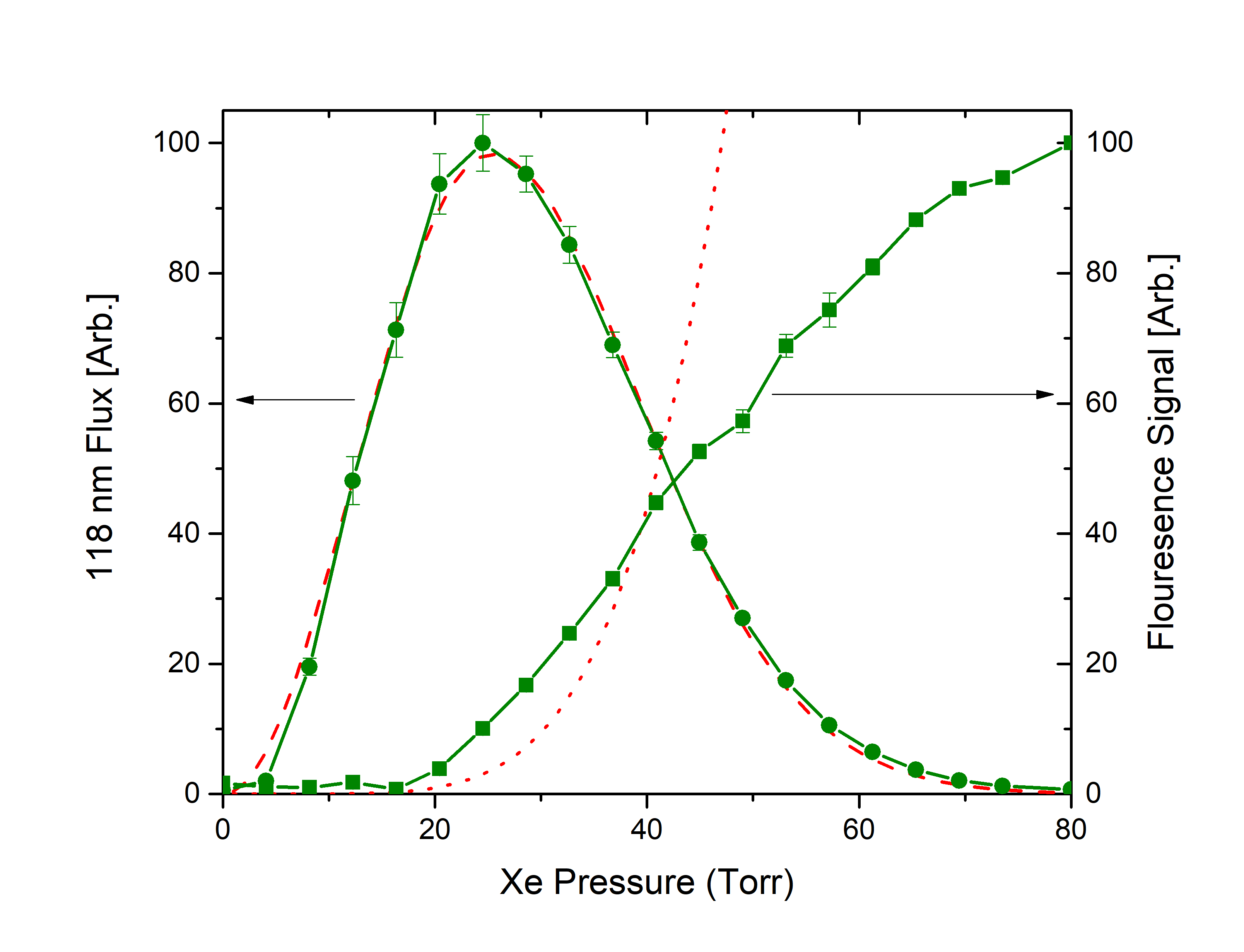}
\caption{\label{fig:ArSim} Comparison of simulated and experimental results for Xe/Ar. Experimental 118 nm flux and fluorescence signal are plotted as green circles and squares respectively. Simulated 118 nm flux and fluorescence signal are plotted as red dashes and dots respectively. The simulated fluorescence curve is independently scaled by eye.
}
\end{figure}

The fitted simulation results for the Xe/Ar case are shown in Fig \ref{fig:ArSim}. The model accurately reproduces the shape of the experimental 118 nm flux. A value of $\alpha_{Xe} = 0.84(1) \times 10^{-37} cm^5$ is extracted from the fit, where the error bar represents only the statistical uncertainty in the fit. We can compare this value of $\alpha_{Xe}$ to one calculated using an oscillator strength $f=0.379$ from Chan \textit{et al.}  \cite{Chan1992} and a collisional broadening coefficient $\beta = 5.6 GHz/Torr$ extracted from the experimental data in Ferrel \textit{et al.}  \cite{Ferrell1987}. The two lines in xenon nearest to the 118 nm light are the Xe ($^2P_{3/2}$)5d $\left[\frac{3}{2}\right]$ line at 119.2 nm and the Xe ($^2P_{3/2} $)7s line at 117 nm. The contribution due to the 117 nm line is negligible since the oscillator strength is a factor of 4 smaller and the collisional line width is a factor of 5 smaller than the 119.2 nm line. The expected value of $\alpha_{Xe}$ is calculated to be $1.18\times 10^{-37} cm^5$, in reasonable agreement with our experimentally determined fit value considering the simple model used.

\begin{figure}[htbp]
\centering
\includegraphics[width=3.5in]{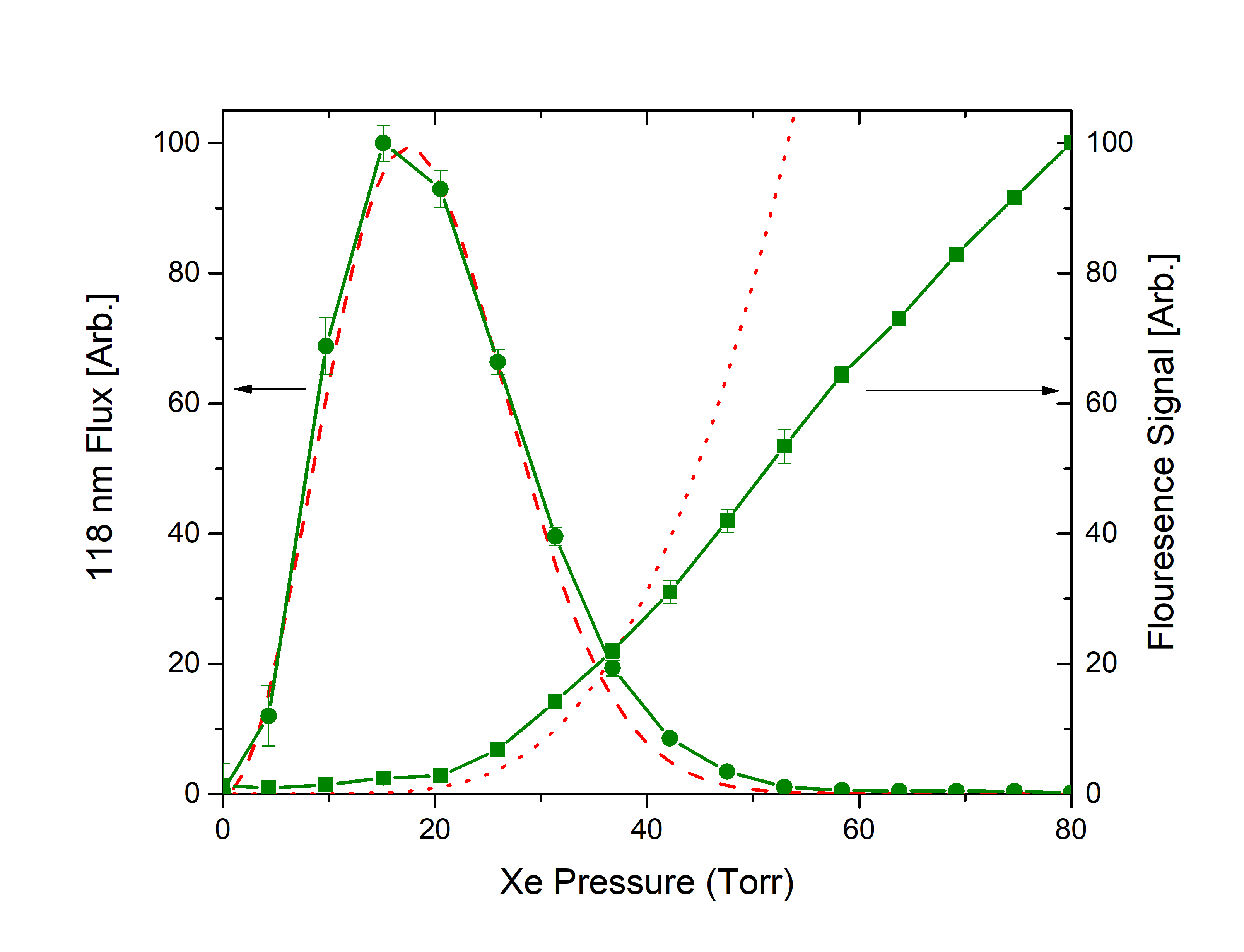}
\caption{\label{fig:KrSim} Comparison of simulated and experimental results for Xe/Kr. Experimental 118 nm flux and fluorescence signal are plotted as green circles and squares respectively. Simulated 118 nm flux and fluorescence signal are plotted as red dashes and dots respectively. The simulated fluorescence curve is independently scaled by eye.
}
\end{figure}

In the case of phase matching with krypton, the Kr ($^2P_{1/2}$)5d $\left[\frac{1}{2}\right]$ resonance line at 116.5 nm also leads to absorption of 118 nm light. This can be accounted for with an additional loss term of

\begin{equation}
\left( \frac{d|A_{118}|}{dz}\right)_{Kr} =- \frac{1}{2} \alpha_{Kr} \mathcal{N}_{Kr}^2 |A_{118}|,
\label{Eq:LossKr}
\end{equation}

\noindent where $\alpha_{Kr}$ is the pressure-broadening coefficient due to krypton and $\mathcal{N}_{Kr}$ is the density of krypton. Since the density of xenon and krypton are related by the phase matching ratio, Eqn. (\ref{Eq:Model2}) can be rewritten as 

\begin{equation}
\frac{d|A_{118}|}{dz} = \frac{1}{2}\chi^{(3)}\mathcal{N}_{Xe}k_{118}|A_{355}|^3 - \frac{1}{2} \alpha_{Xe/Kr} \mathcal{N}_{Xe}^2 |A_{118}|.
\label{Eq:ModelKr}
\end{equation}

\noindent where $\alpha_{Xe/Kr} = \alpha_{Xe} + \alpha_{Kr} R^2$. The results of this model are fit to the experimental Kr results in Fig \ref{fig:KrSim}, with the model reproducing the shape of the experimental curve. From the fit, a value of $\alpha_{Kr} = 4.6(2) \times 10^{-39} cm^5$,  can be compared to an expected value of $\alpha_{Kr} = 7.1 \times 10^{-39} cm^5$ calculated using an oscillator strength $f=0.193$ from Chan \textit{et al.}  \cite{Chan1992} and a collisional broadening coefficient $\beta = 3.0 GHz/Torr$ extracted from the experimental data in Ferrel \textit{et al.}  \cite{Ferrell1987}.

For both Xe/Ar and Xe/Kr, the fluorescence results from the model do not match the experimental results well. However, there is some qualitative agreement. The fluorescence appears to turn on near the pressure where the 118 nm flux starts to turn over. The experimental fluorescence appears to increase linearly while the simulated results appear to increase non-linearly. This may be attributed to uncertainties in the fluorescence measurement. The fluorescence streak is several cm long, and extends beyond the window through which we have optical access to collect fluorescence. We collect only a small portion of the fluorescence streak around the focus.

\section{Discussion and Conclusions}

Third harmonic generation of 118 nm light in a Xe/Ar gas cell is a widely used table top source of VUV radiation. The yield of 118 nm light has been observed to diverge from simple theoretical predictions. The observation of ionized xenon at the focus of the 355 nm beam in the presence of 118 nm light suggests two photon ionization, led by absorption of 118 nm light, is the mechanism limiting the amount of 118 nm light that exits the gas cell. To confirm this mechanism, we developed a model based on pressure-broadened absorption that can be modeled using a simple 1-D simulation. The simulated results are fit to experimental data to extract a pressure-broadening constant, which agrees with values calculated from theory. 

A 118 nm light source could be optimized by considering design options that limit the amount of absorption. Minimizing the amount of xenon the 118 nm light must pass through will increase the yield. This can be accomplished by shortening the gas cell and tightening the focus, but the damage threshold of optical components must be considered. Another option might be using a jet of gas rather than a gas cell. While third harmonic generation has been studied in a pulsed supersonic jet of pure xenon, \cite{KUNG1983, lhuillier1988} jets of phase matched mixtures have not yet been studied. 



\begin{acknowledgments}

We thank E. A. Cornell, M. Murnane, H. C. Kapteyn, D. Adams, A. Jaron-Becker, and A. Becker for useful discussions. This work was supported by the National Science Foundation (PHY-1734006, CHE-1900294) and the Air Force Office of Scientific Research  (FA9550-16-1-0117).

\end{acknowledgments}

\section{Data Availability}

The data that support the findings of this study are available from the corresponding author upon reasonable request.

\bibliography{118paper_v6}

\end{document}